\documentstyle [12pt] {article}

\newcommand{\be}{\begin{equation}}
\newcommand{\ee}{\end{equation}}
\newcommand{\begf}{\begin{figure}}
\newcommand{\ef}{\end{figure}}

\begin{document}
\begin{center}
COMMENT ON THE KAMIOKANDE ATMOSPHERIC NEUTRINO DEFICIT
\footnote{This manuscript is the result of an observation I made publicly
at the {\it XXXth Rencontres de Moriond}, March 12-18, 1995.}\\
{}~\\
David Saltzberg\\
{\it CERN-PPE division, CH-1211, Geneva 23, Switzerland.}\\
31 March 1995\\
\end{center}

\begin{center}
{\it Abstract}\\
\end{center}
\hspace{0.75in}\begin{minipage}[t]{5in}
I describe an attempt to understand the significance
of the atmospheric neutrino deficit observed by the Kamiokande
neutrino detector.  In particular, I am concerned with the
statistical significance quoted for the zenith-angle dependence of the
deficit, which has been cited as evidence
for neutrino flavor oscillations free of systematic uncertainties.\\
\end{minipage}
{}~\\

The Kamiokande neutrino detector collaboration has measured the flux of
electron neutrinos and muon neutrinos produced in cosmic-ray showers in
the Earth's atmosphere [1]. They observe a deficit in the flux of muon
neutrinos (``$\mu$-like events'') relative to the flux of electron neutrinos
(``$e$-like events'') for neutrinos with energies above about 1.3~GeV.   The
Kamiokande collaboration presents two pieces of evidence that their data
are inconsistent with standard-model neutrinos without flavor
oscillations.  First, the ratio of total number
of $\mu$-like to total number of $e$-like events,
$(\mu/e)_{\rm Data}$, divided by the expected ratio, $(\mu/e)_{\rm MC}$, is
inconsistent with unity.  Second, they observe an angular dependence of
this ``ratio of ratios'', $(\mu/e)_{\rm Data}/(\mu/e)_{\rm MC}$.  The latter
effect, if significant,
would be more suggestive, if not
compelling, since it would demonstrate a direct
dependence on the distance to the cosmic ray
shower producing the neutrinos that is free of systematic
uncertainties.  Presumably, the oscillations
of muon neutrinos into another type of neutrino over distances
characteristic of the Earth's radius are being observed.\\

The data are summarized in Figures~3 and~4 of Reference [1].  Figure~3
shows the absolute rate of $e$-like and $\mu$-like events binned by zenith
angle. Figure~4 shows the ``ratio of ratios'' with the same binning.
Although the uncertainties
are all approximately the same size in the raw data presented in
Figure~3,  the uncertainties vary significantly from bin to bin in Figure~4.
Examining the numbers shows that the Kamiokande group has propagated
uncertainties estimated from the data points themselves to
arrive at the uncertainty on each point in Figure~4.  As a result
points which might be low merely due to a fluctuation, will have
uncertainty estimates that are too low.  While such estimates for the
uncertainties   may be the best available for data for which there is no
model, they are inappropriate for data which are subsequently compared
to a particular hypothesis.  For testing a hypothesis,
one would prefer to use
uncertainties based on the {\it expected} rather than {\it observed} rates.\\

In this reanalysis, the raw count rates are used directly to
test if the atmospheric neutrino deficit has a zenith-angle dependence
inconsistent with any flat distribution.
A fit is performed to
minimize the $\chi^2$ recommended [2] for rates following
a Poisson distribution:
\begin{equation}
\chi^2 = \sum_{\ell=e,\mu}\ \sum_{i=1}^{5} \
[2 (N_{i}^{\rm exp}-N_{i}^{\rm obs}) \
+ \ 2 N_{i}^{\rm obs} \ {\rm ln}(N_{i}^{\rm obs}/N_{i}^{\rm exp})],
\end{equation}
where the sums are over the five bins in both the electron- and
muon-neutrino distributions and
the $N_{i}^{\rm exp}$ are given by:
\begin{center}
\begin{tabular}{lll}
$N_{i}^{\rm exp}(e)$ & = & $\alpha$ $N_{i}^{\rm exp-SM}(e)$\\
$N_{i}^{\rm exp}(\mu)$ & = & $\alpha \beta N_{i}^{\rm exp-SM}(\mu)$.\\
\end{tabular}
\end{center}
The parameters $\alpha$ and $\beta$ are chosen to minimize $\chi^{2}$.
The parameter $\alpha$ absorbs an overall
normalization; the
parameter
$\beta$ is the
correction factor to the ``ratio of ratios'' sought by Kamiokande.\\

The fit to the Kamiokande data yields $\beta=0.56^{+0.08}_{-0.07}$(stat.),
in good agreement with the Kamiokande
result of $0.57^{+0.08}_{-0.07}$(stat.), from a different method.
The calculated $\chi^2/{\rm d.o.f.}$
is 15.4/8=1.93.  The corresponding
confidence level with which a flat distribution is
excluded by the data is 94.8\%.  This is
less than the ``conservative'' 99\% claimed in the Kamiokande paper
and much less than the confidence level conventionally required to claim
a signal.  As a check,
the same fits are done to 5000 Monte Carlo experiments.  The returned values
of $\alpha$ and $\beta$ follow the expected distributions.  The number of
experiments with $\chi^{2}>15.4$ is 5.8\%, in agreement with
the extracted 94.8\% confidence level.\\


In conclusion,  the observed deviation of the Kamiokande ``ratio of
ratios'' from flatness is expected in  1 in 20 such experiments.  It is
noted that the uncertainties on  the data points in Figure~4 of
Reference [1] are misleading.\\

I thank S. Eno, H. Frisch and K. Ragan for helpful discussions.\\
{}~\\
\hspace{0in}[1] Y. Fukuda {\it et al.}, Phys. Lett. {\bf B335} (1994) 237.\\
\hspace{0in}[2] S. Baker and R. D. Cousins, Nucl. Instr. and Meth., {\bf 221}
(1984) 437.
\end{document}